\documentclass{article}
\usepackage{slashed,amsmath,amssymb,enumitem}
\usepackage[papersize={8.5in,11in}]{geometry}
\begin{document}
\title{A Composite Model of Quarks and Bosons}
\author{J. W. Moffat\\~\\
Perimeter Institute for Theoretical Physics, Waterloo, Ontario N2L 2Y5, Canada\\
and\\
Department of Physics and Astronomy, University of Waterloo,\\
Waterloo, Ontario N2L 3G1, Canada}
\maketitle
\begin{abstract}%
A composite model of quarks and bosons is proposed in which a spin $1/2$ isospin doublet $\psi$ is the basic building block of quarks and bosons in the standard model. The $\psi$ has two components $v$ and $w$ with charges $Q=\frac{1}{3}e$ and $Q=0$, respectively, that combine to form the three generations of colored quark flavors.  A strong force described by a triplet of massless gluons bind the constituents called geminis. The confining constituent non-Abelian $SU(2)_C$ field theory is called constituent dynamics with a confining energy scale $\Lambda_{CD}$. The constituent dynamics condensate $\langle\bar{v}v+\bar{w}w\rangle\neq 0$ spontaneously breaks the electroweak symmetry $SU(2)_L\times U(1)_Y\rightarrow U(1)_{\rm EM}$ and a triplet of Nambu-Goldstone bosons make the gauge bosons $W^{\pm}$ and $Z^0$ massive, while retaining a massless photon. A global custodial $SU(2)_L\times SU(2)_R$ symmetry guarantees that the symmetry breaking in the weak interaction sector agrees with electroweak data. The non-Abelian $SU(2)_C$ color dynamics satisfies asymptotic freedom, which resolves the gauge and Higgs mass hierarchy problems and makes the model ultraviolet complete. The composite constituent dynamics model can realize a $SU(3)_C\times SU(2)_L\times U(1)_Y$ electroweak and strong interaction model that satisfies the naturalness principle. The three generations of colorless quarks $\alpha$ and $\beta$ with charges $Q=+1e$ and $Q=0$, respectively, which are predicted to exist in the composite model can form bound states which can be identified with the spectrum of exotic mesons.
\end{abstract}



\section{Introduction}

The LHC collaboration has discovered a Higgs-like boson at a mass $125-126$ GeV~\cite{ATLAS,CMS}. This renews the problems of the fine-tuning of the Higgs boson mass and the gauge hierarchy, because of the quadratic divergences in the radiative corrections to the scalar boson bare mass~\cite{Veltman,Hamada,Jones} and that $v << \Lambda$ where $v=246$ GeV and $\Lambda$ is an energy cutoff. The Higgs boson self-energy $\delta m_H^2\sim\Lambda^2/16\pi^2$ leads to large corrections for an arbitrarily large $\Lambda$. If an $\overline {MS}$ subtraction is performed on the bare scalar field sector Lagrangian, the renormalized Lagrangian can be formulated in the renormalization group language, so that $\delta m_H^2\propto M^2\ln(M^2/\mu^2)$ where $M$ is a renormalized mass and $\mu$ is the renormalization group mass parameter.  If no heavy mass particles exist above the top quark mass and there are no large energy scales such as the GUT scale at $\sim 10^{15}$ GeV, then the standard model (SM) can be free of fine-tuning and satisfy the ``naturalness'' principle~\cite{Wilson,tHooft}. So far, the LHC has not detected any new physics beyond the standard model. However, if new heavy particles exist, then a fine-tuning naturalness problem emerges, for the Higgs boson mass radiative correction gives $\delta m_H^2\sim \lambda_{tH}M^2/16\pi^2$ for a heavy particle of mass $M$ where $\lambda_{tH}$ is the Higgs-top quark coupling constant~\cite{Barbieri}. Moreover, the electroweak (EW) vacuum can be critically metastable at a high energy scale~\cite{Branchina}. Because the SM has $\sim$ 20 free parameters and the potential exists for new physics to be discovered at the LHC and at future accelerators, the SM can only be considered an incomplete theory subject to hierarchy problems and fine-tuning. 

QCD based on confined colored quarks and colored gluons can be considered an example of a fundamental gauge theory~\cite{Gellmann,Zweig,Gellmann2}. The QCD Lagrangian is chirally invariant when the quark masses are zero, so that at the classical level the theory is conformally scale invariant. The only dimensionless constant is the strong coupling constant $\alpha_s$. The color charge of the quarks is screened at high energies and this color charge screening produces asymptotic freedom~\cite{Gross,Politzer}. QCD quantum fluctuations are damped at high momenta, so that there is no need to introduce a high momentum cutoff. The quark masses become non-zero in the low energy infrared limit when the gluon confining interaction becomes strong, producing an explanation for the rich hadronic spectrum of particles. The experimentally determined QCD energy scale $\Lambda_{\rm QCD}\sim 200$ MeV removes the scale invariance of the quantized theory through chiral condensates, which break the chiral symmetry of the Lagrangian. The theory is ultraviolate (UV) complete making it a fundamental gauge field theory. The scalar sector of the EW interactions in the SM prevents the theory from possessing asymptotic freedom and a damping at high energies. A fundamental SM theory would treat the  spontaneous symmetry breaking Higgs mechanism~\cite{Anderson,Englert,Higgs,Kibble,Weinberg,Salam} as an effective low energy, infrared description of particle interactions.

The existence of isotopes in nuclear physics, the proliferation of elements in Mendeleev's periodic table and the systematics in the organization of the table, strongly suggested that a substructure existed, leading to the build up of the elements by the more fundamental electrons, protons and nuclei. The proliferation of baryons and mesons (hadrons) as ``elementary" particles and their systematics, pointed the way to the quark substructure of nucleons in a replay of the motivation for the existence of composite atoms based on Mendeleev's table. The leptons and neutrinos do not experience strong interactions, while the neutrinos participate exclusively in the weak interactions. The QCD $SU(3)_C$ color gauge theory was formulated to explain the confinement of quarks and the substructure underlying hadrons.

Early composite models of quarks and bosons postulated that the leptons and the force carrying bosons $W, Z,\gamma$ and the gluons were composite particles~\cite{Salam2,Neeman,Harari,Shupe,Adler,Terazawa,Greenberg,Fritzsch,Abbott, Barbieri2,Calmet,Fritzsch2}. The model proposed in the following retains the point-like nature of the leptons, massless gluons and the photon \footnote[1]{Earlier proposals also made leptons, gluons and the photon composite particles~\cite{Harari,Shupe} and this remains a possible option.}. The proposed composite model is expected to produce the final primary building blocks of matter, for the basic spin 1/2 $SU(2)$ constituents called geminis cannot be reduced by group theoretical considerations to simpler substructures. 

The LHC has not currently detected any composite substructure of the standard model fermions and bosons. Arguments make us believe that possible substructures will only be observed at very short distances and large momenta, which are expected to occur for distances less than $10^{-16}\,{\rm cm}$. Perhaps, at the maximum energy of 14 TeV at the LHC, we can reach distances less than $10^{-16}\,{\rm cm}$ and with the future ILC, it will be possible to probe distances of order $10^{-18}\,{\rm cm}$. 

One prediction of our composite model is the existence of colorless quark states $\alpha$ and $\beta$ with charges  $Q=+1e$ and $Q=0$, respectively. The bound states of these quarks produce the $W^{\pm}, Z^0$ and the Higgs boson $H^0$. We predict that they also form bound states of observed exotic mesons, such as the $X(3872)$~\cite{LHCb,EPJ} with the quantum numbers $J^{PC}=1^{++}$.

\section{Composite Model}

The most economical model to consider for the basic building blocks of matter is the spin 1/2 isodoublet $\psi=\begin{pmatrix}v\\w\end{pmatrix}$. We call the constituent component particles $v$ and $w$ geminis\footnote[2]{Gemini is the Latin word for twin. They are not identical twins. Earlier names for the constituent particles are preon~\cite{Salam2}, rishon~\cite{Harari}, quip~\cite{Shupe} and haplon~\cite{Fritzsch}.}. The charges of the $v$ and $w$ are $Q=\frac{1}{3}e$ and $Q=0$, respectively~\cite{Harari,Shupe}. The charge $Q$ is expressed as the Gell-mann-Nishijima formula:
$Q=I_3+\frac{1}{2}Y$ where $I_3$ and $Y$ are the third component of isospin and strong hypercharge, respectively. We then have $I_{3v}=1/2$, $I_{3w}=-1/2$, $Y_v=1/3$ and $Y_w=1$. The $Q=2/3$ up quark flavor is formed from the three color combinations: 
\begin{equation}
\label{upquarkcolor}
u(vvw)_R, u(vwv)_B, u(wvv)_G,
\end{equation}
while the $Q=-1/3$ down quark flavor is formed from the three color combinations: 
\begin{equation}
\label{downquarkcolor}
d({\bar v}{\bar w}{\bar w})_R, d({\bar w}{\bar v}{\bar w})_B, d({\bar w},{\bar w}, {\bar v})_G.
\end{equation}
Here, the $v$ and $w$ are treated as non-commuting spinor matrices~\cite{Harari,Shupe}. The up and down quarks degenerate color combinations are for strangeness number $S=0$. The two additional up and down quark type generations of flavors have the same color combinations of $v,w$, but they are distinguished by the quantum numbers $S=-1$ for the strange quark $s$, C=+1, for the charm quark $c$, $B'=-1$ for the bottom quark $b$, and $T=+1$ for the top quark $t$. The charge $Q$ now becomes
\begin{equation}
Q=I_3+\frac{1}{2}(B+S+C+B'+T),
\end{equation}
where $B$ denotes the baryon number. The two combinations $vvv$ and $www$ are additional quarks with charges $Q=+1e$ and $Q=0$, respectively. They do not carry $SU(3)_C$ color charge and we call the two colorless quarks $\alpha(vvv)$ and $\beta(\bar{w}\bar{w}\bar{w})$. The $\alpha$ and $\beta$ also come in three generations $\alpha_1,\alpha_2,\alpha_3$ and $\beta_1,\beta_2,\beta_3$. 

We treat the leptons, $e,\mu,\tau,\nu_e,\nu_\mu,\nu_\tau$ as point particles.\footnote[2]{Other authors also make the leptons composites, a possible option in the model. The combinations $vvv$ and $www$ are identified with $e^+$ and $\nu_e$.}. As in the case of the octet of gluons in $SU(3)_C$ that confines the SM quarks the $SU(2)_C$ gluons are electrically neutral and massless. We expect that for the consituent gemini electromagnetic current, we have
\begin{equation}
J^\mu_{{\rm em}}=\frac{1}{\sqrt{2}}({\bar v}_L\gamma^\mu v_L+{\bar v}_R\gamma^\mu v_R).
\end{equation}

The spin 1 gauge bosons $W^{\pm}$ and $Z^0$ are formed from combinations of the $\alpha$ and $\beta$ quarks: $W^+(\alpha\bar\beta), W^-(\bar\alpha\beta)$ and $Z^0(\alpha\bar\alpha-\beta\bar\beta)$, while the composite Higgs boson is formed from $ H^0(\alpha\bar\alpha+\beta\bar\beta)$. The neutral left and right components of the gauge boson $W^3_\mu$ are given by $W^3_{\mu L,R}=\frac{1}{\sqrt{2}}({\alpha}_{L,R}\gamma_\mu\bar\alpha_{L,R}+{\beta}_{L,R}\gamma_\mu \bar\beta_{L,R})$. The neutral singlet gauge boson $B_\mu$ is orthogonal to $W^3_{\mu L}$ and $W^3_{\mu R}$: $B_\mu=\frac{1}{2}({\alpha}_L\gamma_\mu \bar\alpha_L+{\alpha}_R\gamma_\mu \bar\alpha_R-{\beta}_L\gamma_\mu\bar\beta_L-{\beta}_R\gamma_\mu\bar\beta_R)$.

We must now construct a dynamical model of the constituent $v$ and $w$ and the gluons that bind the constituents in the quarks and the quarks in the hadrons. The $v$ and $w$ have two colors belonging to the representation $\underline 2$ of $SU(2)_C$ and we have $\underline 2\times \underline 2^*=\underline 1+\underline 3$. The triplet $\underline 3$ of gluons couple to the two colors red and yellow, $v_R,v_Y$ and $w_R,w_Y$. The non-Abelian $SU(2)_C$ gluons and the constituent dynamics (CD) confine the colored geminis inside the quarks, which in turn carry the the three $SU(3)_C$ red, blue and green color charges formed from the combinations (\ref{upquarkcolor}) and (\ref{downquarkcolor}) and the two additional flavor generations of colored $s,c,b$ and $t$ quarks. As in the SM the hadrons are formed from the singlet (white) color combinations of quarks. 
We adopt the $SU(n)_C$ color charge Lagrangian:
\begin{equation}
\label{CDLagrangian}
{\cal L}_{C}={\bar \chi}_i(i\slashed\partial-m_{\chi_i})\chi_i-ig_C{\bar\chi}_i(\gamma^\mu T_aG^a_\mu)\chi_i-\frac{1}{4}G^a_{\mu\nu}G^{\mu\nu}_a,
\end{equation}
where the Dirac spinors $\chi_i$ have internal flavor and color charge components, $\slashed\partial$ denotes $\gamma^\mu\partial_\mu$, $T_a$ are the generators of the group $SU(n)_C$ and $g_{\rm C}$ is the constituent dynamic coupling constant. We have
\begin{equation}
G^a_{\mu\nu}=\partial_\mu G_\nu^a-\partial_\nu G^a_\mu -f_{abc}G^b_\mu G^c_\nu.
\end{equation}
The required gauge invariance of ${\cal L}_{C}$ can be achieved provided we have the gauge transformation:
\begin{equation}
G_{a\mu}\rightarrow G_{a\mu} - \frac{1}{g_{\rm C}}\partial_\mu\theta_a -f_{abc}\theta_b G_\mu^c,
\end{equation}
where $\theta$ is the phase in the transformation:
\begin{equation}
\chi_i(x)\rightarrow U\chi_i(x)=\exp(i\theta_a(x)T_a)\chi_i(x),
\end{equation}
and $U$ is an $n\times n$ unitary matrix.

For vanishing masses $m_{\chi_ i}=0$, the Lagrangian (\ref{CDLagrangian}) with $\chi_{iL},\chi_{iR}$ for the consituent Dirac fields posseses a scale invariant and chiral symmetry. The scale invariance is broken by the constituent confinement energy scale $\Lambda_{CD}$. We can anticipate that the CD confining $SU(2)_C$ coupling constant $\alpha_C=g^2_{\rm C}/4\pi$ will be screened and lead to asymptotic freedom. This will cause a damping of high energy momentum processes and lead to a UV complete model, analogous to the flavor color screening and asymptotic freedom of QCD. 

For the gauge color group $SU(N)_C$ the beta function in lowest order perturbation theory is given by
\begin{equation}
\beta(g_C)=-\frac{g_C^3}{48\pi^2}\biggl(11C_2-\frac{4}{3}N_fC(R)\biggr),
\end{equation}
where $C_2$ is the quadratic Casimir coefficient of $SU(N)_C$ and $C(R)$ is the Casimir invariant $Tr(T^a_RT^b_R)=C(R)\delta^{ab}$ for the group generators $T_R^{ab}$ of the Lie alegbra in the representation $R$. For gluons in the adjoint representation of $SU(N)_C$, we have $C_2=N_C$ (where $N_C$ is the number of colors) and for fermions in the fundamental representation: $C(R)=1/2$.  For $SU(3)_C$ in QCD, $N_C=3$ and for the constituent dynamics, $N_C=2$. 
For the color gauge group $SU(2)_C$ the running CD constituent coupling constant $\alpha_C(q^2)$ is given by
\begin{equation}
\label{runningalpha}
\alpha_C(q^2)=\frac{\alpha_C(\mu^2)}{1+\frac{\alpha_C(\mu^2)}{12\pi}(22-2N_{cf})\ln(q^2/\mu^2)},
\end{equation}
where $\mu$ is a renormalization group flow mass and the number of constituent flavors $N_{cf}=2$. For sufficiently low $q^2$, the effective coupling will become large. This will occur at the constituent confinement scale $\Lambda_{CD}$:
\begin{equation}
\Lambda_{CD}^2=\mu^2\exp\biggl[-\frac{12\pi}{(22-2N_{cf})\alpha_C(\mu^2)}\biggr].
\end{equation}
We can now write (\ref{runningalpha}) in the form:
\begin{equation}
\label{alpha}
\alpha_C(q^2)=\frac{12\pi}{(22-2N_{cf})\ln(q^2/\Lambda_{CD}^2)}.
\end{equation}
For $q^2$ values much larger than $\Lambda_{CD}^2$, the effective coupling is small and this allows for a perturbative description of the geminis and gluons. For $q^2\sim\Lambda_{CD}^2$, we cannot use perturbation theory and the geminis arrange themselves into strongly bound SM quarks. The parameter $\Lambda_{CD}$ which enters the model through renormalization is to be determned by experiment.

The LHC experiments have shown that quarks and leptons are point-like down to a distance scale $< 10^{-16}$ cm, or less than about $10^{-3}$ of the proton diameter. If the constituent particle is confined to a box with a side of order $10^{-16}$ cm, then the momentum uncertainty following from the uncertainty principle, $\Delta x\Delta p\geq \hbar/2$, is $\Delta p\sim$ 200 GeV, which is about $5\times 10^4$ times bigger than the rest mass momentum of an up quark. Thus, the composite quark model suggests that the lighter composite quarks have a large kinetic energy, because the momentum uncertainty $\Delta p$ of the constituent particles will be greater than the composite quark momentum. This apparent paradox is resolved by assuming a large binding energy $E_B\sim\Lambda_{\rm CD}$ in the GeV range of energy between the constitutent particles, and it indicates that the dynamics of the light composite quarks is relativistic.

In the SM the mass parameter $\mu$ and the coupling constant $\lambda$ in the minimum:
\begin{equation}
\phi^\dagger\phi\equiv\frac{v_{\rm EW}^2}{2}=-\frac{\mu^2}{2\lambda},
\end{equation}
of the Higgs potential:
\begin{equation}
V(\phi)=\mu^2\phi^\dagger\phi+\lambda(\phi^\dagger\phi)^2,
\end{equation}
are after quantization not stable against quantum corrections. The mass parameter $\mu$ and so the electroweak scale $v_{\rm EW}\sim$ 246 GeV generate quadratically divergent quantum corrections for the Higgs mass $m_H$ in contrast to the other parameters which only aquire logarithmic corrections. The parameter $\lambda$ runs according to the renormalization group equation, dominated by the top quark Yukawa coupling constant $\lambda_{\rm tH}$. It runs to either a Landau pole or to zero at some high energy scale $\Lambda$. The SM then suffers the {\it gauge hierarchy problem}, $v_{\rm EW} <<\Lambda$, where $\Lambda$ is an energy scale at which the quantum corrections are significant and $\Lambda$ can be extended to the Planck energy $\Lambda_{\rm PL}\sim 10^{19}$ GeV. The fine-tuning or naturalness problem makes the SM with the Higgs EW symmetry breaking a low energy, effective infrared phenomenological model and not a fundamental theory. 

The gauge hierarchy problem can be solved in a natural way as in standard QCD through the asymptotic freedom~\cite{Gross,Politzer} associated with the exact non-Abelian $SU(2)_C$ symmetry. A UV fixed point at which $\alpha_C$ tends to zero according to (\ref{alpha}):
\begin{equation}
\alpha_C(q^2)\propto \frac{1}{\ln(q^2/\Lambda^2_{\rm CD})},
\end{equation}
invokes a natural cutoff for quantum corrections, which are kept under control. Because $\alpha_C(\Lambda_{\rm PL}) << 1$ at the Planck energy scale $\Lambda_{\rm PL}$ a large exponential suppression is invoked:
\begin{equation}
\Lambda^2_{\rm CD}\sim\exp\biggl(-\frac{1}{\alpha_C(\Lambda_{\rm PL}^2)}\biggr)\Lambda^2_{\rm PL},
\end{equation}
which solves the gauge hierarchy problem. Our composite model of quarks and bosons will describe a theory which is free of the SM fine-tuning problems and is UV complete.

\section{Electroweak Symmetry Breaking}

The $SU(2)_C$ color fields attraction at small momentum $q^2 < \Lambda_{\rm CD}$ will alter the structure of the vacuum. Let us assume that the constituent doublet $\psi=\begin{pmatrix}v\\w\end{pmatrix}$ engages in electroweak as well as strong interactions. A chiral condensate: 
\begin{equation}
\label{condensate}
\langle\bar{v}_L v_R+\bar{w}_Lw_R\rangle
\end{equation}
will be generated, which spontaneously breaks the chiral symmetry $SU(2)_R\times SU(2)_L$ of the Lagrangian\begin{footnote}{A similar spontaneous symmetry mechanism is introduced in QCD and technicolor~\cite{Susskind}}\end{footnote}. We turn on the $SU(2)_L\times U(1)_Y$ electroweak interaction without the scalar fields usually introduced to give masses to the $W_\mu^a$ associated with $SU(2)_L$  and the $B_\mu$ associated with weak hypercharge. Now three massless Nambu-Goldstone bosons $\pi^a$ play the role of the massless scalar fields and become the longitudinal components of the massive $W_\mu^{\pm}$ and $Z_\mu^0$.  

When the constituent dynamics is coupled to EW gauge dynamics, the $CD$ condenstate (\ref{condensate}) spontaneously breaks the EW gauge dynamics. The Goldstone theorem says that there will be a massles spin 0 boson for each broken generator. This means that there are three massless Nambu-Goldstone bosons forming an isotriplet. The constituent particle decay constant $f_{\rm C}$ is defined by
\begin{equation}
\langle 0\vert J^{\pm}_\mu\vert \pi^{\pm}\rangle=f^{\pm}_{\rm C}q^\mu,
\end{equation}
and
\begin{equation}
\langle 0\vert J^Y_\mu\vert \pi^0\rangle=f^0_{\rm C}q^\mu,
\end{equation}
where $J^{\pm}_\mu$ are two $SU(2)_L$ currents and $J^Y_\mu$ is the hypercharge current. Because of the conservation of electromagnetic charge, $f^+_{\rm C}=f^-_{\rm C}$, and due to the custodial symmetry $SU(2)_{\rm cust}$ we have $f^{\pm}=f^0=f_{\rm C}$. The $W_\mu^{\pm}$ now have a mass:
\begin{equation}
M^2_W=\frac{1}{4}g^2f^2_{\rm C},
\end{equation}
where $f_{\rm C}\sim$ 246 GeV.

For the neutral gauge bosons, we need a mass matrix to describe the $2\times 2$ matrix for the Abelian $B_\mu$ and the $W^3_\mu$ component of the $W^a_\mu$ boson. We have
\begin{equation}
M^2=\frac{1}{4}\begin{pmatrix}
g^{2} & gg'\\
gg' & g^{'2}
\end{pmatrix}f^2_{\rm C},
\end{equation}
where $g$ and $g'$ are the $SU(2)_L$ and $U(1)_Y$ coupling constants, respectively. The eigenvalues of this matrix are given by
\begin{equation}
M_\gamma^2=0,
$$ $$
M^2_Z=\frac{1}{4}(g^2+g^{'2})f_{\rm C}^2,
\end{equation}
where $M_\gamma$ and $M_Z$ denote the photon and $Z$ masses. We have 
\begin{equation}
\frac{M_W}{M_Z}=\frac{g}{(g^2+g^{'2})^{1/2}}=\cos\theta_w,
\end{equation}
where $\theta_w$ is the weak mixing angle. We obtain the standard tree level result:
\begin{equation}
\rho=\frac{M_W^2}{M_Z^2\cos^2\theta_w}=1.
\end{equation}
According to the symmetry group pattern, the spectrum of composite quarks is determined. The quark masses are directly proportional to $\Lambda_{\rm CD}$.

As in QCD in which the masses of hadrons are generated dynamically, the masses of the quarks and leptons in our composite model will be generated dynamically. In contrast, in the SM the masses of the quarks and leptons are generated by their Yukawa couplings to the Higgs boson. In the SM the masses of fermions are parameterized by their associated coupling constants. {\it This fits the data but no explanation is given for the magnitudes of the particle masses.} 

\section{Exotic Mesons}

There has been mounting evidence for the existence of exotic mesons, which it is speculated represent a new form of matter. One such meson is $X(3872)$~\cite{LHCb,EPJ} with a mass $m_X=3.872$ GeV.  Recently its quantum numbers have been determined by the LHCb collaboration~\cite{LHCb} to be $J^{PC}=1^{++}$, favoring an exotic explanation. We venture to predict that the $X$ meson state is formed from an isoscalar second generation $\alpha(vvv)$ and $\beta(\bar{w}\bar{w}\bar{w})$ neutral quark bound state $X(\alpha_2\bar\alpha_2+\beta_2\bar\beta_2)$. We name the $X$ meson $X_{\alpha\beta}$. It will be observed through the decays $B^+\rightarrow X_{\alpha\beta}K^+$, $X_{\alpha\beta}\rightarrow\pi^+\pi^-J/\psi$ and $J/\psi\rightarrow\mu^+\mu^-$. Since the Higgs boson with $J^{PC}=0^{++}$ is composed of the $SU(3)_C$ third generation of corlorless spin $1/2$ quarks, $H^0(\alpha_3\bar\alpha_3+\beta_3\bar\beta_3)$, then we may consider the Higgs boson as an exotic meson resonance with a mass $m_H\sim 125-126$ GeV.

We predict that our composite model is responsible for a rich spectrum of exotic mesons composed of the three generations of $\alpha$ and $\beta$ quarks that complements the standard QCD spectrum of charmonium and bottomonium states. Further work will investigate the nature of the predicted exotic meson spectrum.

\section{Fermion Masses}

A general approach to determining the quark masses is the dynamically broken $SU(2)_L\times SU(2)_R$ chiral flavor symmetry for the constituent color $SU(2)_C$ theory. It is assumed that the bare constituent masses are zero while the dynamical masses $m_c$ are non-zero. The propagator for the constituent flavors $v,w$ is given by~\cite{Jackiw,Pagels,PagelsStokar}:
\begin{equation}
S^{-1}(q^2)=\slashed{q}-\Sigma_c(q^2),
\end{equation}
where the constituent particle self-energy $\Sigma_c(q^2)$ obeys a Bethe-Salpeter equation. An approximate solution to $\Sigma_c(q^2)$ has the asymptotic behavior in the Landau gauge as $-q^2\rightarrow\infty$ ~\cite{PagelsStokar}:
\begin{equation}
\Sigma_c(q^2)\sim \frac{4m_c^2}{q^2}\ln^\gamma\biggl(\frac{-q^2}{\mu^2}\biggr),
\end{equation} 
where  $\gamma=12/(22-2N_{cf})$, $N_{cf}=2$ and $m_c$ is the dynamical constituent mass. 

The quark and lepton masses can be generated by a one-loop self-energy diagram. For the quarks it involves the $\psi$ doublet $v,w$ and the gauge boson and CD gluon loops. We assume that the bare quark and lepton masses are zero. In the Landau gauge, we have for the lepton self-energy:
\begin{equation}
\Sigma_{\rm lept}(q^2)=-\frac{3ig_{\rm lept}^2}{(2\pi)^4}\int d^4k\frac{1}{(k-q)^2-\Lambda_{\rm lept}^2}\frac{\Sigma_{\rm lept}(k^2)}{k^2-\Sigma^2_{\rm lept}(k^2)},
\end{equation}
where $g_{\rm lept}$ is the constituent particle-lepton coupling constant, $\Lambda_{\rm lept}$ is the lepton energy scale constant that breaks the conformal, chiral symmetry of the lepton sector and $\Sigma_{\rm lept}$ denotes the lepton self-energy.  The quark masses are determined by the one-loop self-energy diagram:
\begin{equation}
\Sigma_q(q^2)=-\frac{3ig_{\rm C}^2}{(2\pi)^4}\int d^4k\frac{1}{(k-q)^2-\Lambda_{\rm CD}^2}\frac{\Sigma_{\rm c}(k^2)}{k^2-\Sigma^2_{\rm c}(k^2)}.
\end{equation}

After a Wick rotation, we can obtain approximate expressions for the lepton and quark masses~\cite{Smetana}:
\begin{equation}
m_{\rm lept}\sim\Sigma_{\rm lept}(0)=\frac{3g^2_{\rm lept}}{8\pi^2\Lambda^2_{\rm lept}}\int^{\Lambda^2_{\rm lept}}_0 dy\frac{y\Sigma_{\rm lept}(y)}{y+\Sigma^2_{\rm lept}(y)},
\end{equation}
and
\begin{equation}
m_q\sim\Sigma_q(0)=\frac{3g^2_{\rm C}}{8\pi^2\Lambda^2_{\rm CD}}\int^{\Lambda^2_{\rm CD}}_0 dy\frac{y\Sigma_c(y)}{y+\Sigma^2_c(y)}. 
\end{equation}
The Pagels-Stokar formula~\cite{PagelsStokar} gives the approximate result for the $f_{\rm C}$ decay constant:
\begin{equation}
f^2_{\rm C}\sim\frac{N_{cf}}{8\pi^2}\int^{\Lambda^2_{\rm CD}}_0 dy\frac{y\Sigma^2_c(y)}{(y+\Sigma^2_c(y))^2}.
\end{equation}

\section{Conclusions}

The economical model proposed here suggests that the final basic building block of matter is the two-component doublet $\psi=\begin{pmatrix}v\\w\end{pmatrix}$.
The concept of three generations of colored quark flavors appears naturally. Since the next lower level of group structure is the Abelian $U(1)$, we can 
speculate {\it that the basic constituents of matter based only on spin 1/2 fermions can no longer be reduced to a further substructure}.  The model has a scale invariant chiral symmetry for the massless constituents $v$ and $w$ in analogy to the chiral symmetry of QCD. The scale invariance is broken by the confining energy scale $\Lambda_{CD}$. The condensate $\langle \bar{v}v+\bar{w}w\rangle\neq 0$ spontaneously breaks the electroweak symmetry, $SU(2)_L\times U(1)_Y\rightarrow U(1)_{\rm EM}$, and gives masses to the $W^{\pm}$ and $Z^0$ bosons keeping the photon massless.  The masses of the quarks are generated by the $SU(2)_{\rm C}$ confining dynamics with the triplet of gluons interacting with the $v$ and $w$ constituent geminis. The masses of the quarks and the point-like leptons are determined by the particle quantum field dynamics involving self-energy integral equations. The gemini-gluon confining force satisfies an $SU(2)_C$ color charge $\alpha_C$ screening and asymptotic safety, leading to a UV complete gauge theory and a resolution of the Higgs mass and gauge hierarchy problems. The standard theory of spontaneous symmetry breaking and the generation of fermion masses from the Yukawa fermion-Higgs Lagrangian with $\langle 0\vert\phi_H\vert 0\rangle\neq 0$ is treated as an effective infrared low energy model. The composite model removes the troublesome scalar Higgs mass and gauge hierarchy problems allowing for a model of strong and weak interactions that satisfies the principle of naturalness.

The model predicts the existence of three generations of $SU(3)_C$ colorless spin $1/2$ quarks $\alpha_1, \alpha_2,\alpha_3$ and $\beta_1,\beta_2,\beta_3$, which can generate a spectrum of exotic mesons~\cite{EPJ}. The $X(3872)$ narrow meson resonance with quantum numbers $J^{PC}=0^{++}$ and $I=0$~\cite{LHCb} would be composed of the second generation of $\alpha_2,\bar\alpha_2$ and $\beta_2\bar\beta_2$ quarks with a mass in the charmonium meson energy range $\sim$ 3.8 GeV. 

\section{Acknowledgements}

The John Templeton Foundation is thanked for its generous support of
this research. The research was also supported by the Perimeter
Institute for Theoretical Physics. The Perimeter Institute was
supported by the Government of Canada through Industry Canada and by
the Province of Ontario through the Ministry of Economic Development
and Innovation. We also thank Martin Green, Carlos Tamarit, Wolfgang Altmannshofer and Viktor Toth for helpful discussions and comments.

\end{document}